\documentclass{aa}

\usepackage{graphicx}
\usepackage{color}
\usepackage{float}
\usepackage{wrapfig}
\usepackage{subcaption}
\usepackage{booktabs}
\usepackage{txfonts}
\usepackage[colorlinks=true,allcolors=blue]{hyperref}
\usepackage{booktabs}
\usepackage{euscript}
\usepackage{mathrsfs}

\begin{document}

\title{Temporal variations of solar inertial mode parameters\\ from GONG (2002--2024) and HMI (2010--2024)}
\subtitle{Rossby modes ($3 \leq m \leq 16 $) and $m=1$ high-latitude mode}

\author{
B Lekshmi \inst{\ref{mps},\ref{nso}}
\and
Zhi-Chao Liang \inst{\ref{mps}}
\and
Laurent Gizon \inst{\ref{mps},\ref{gottingen},\ref{NYUAD},\thanks{Corresponding author: gizon@mps.mpg.de}}
\and
Jordan Philidet\inst{\ref{ObsPM}}
\and
Kiran Jain \inst{\ref{nso}}
}

\institute{
Max-Planck-Institut f\"ur Sonnensystemforschung, Justus-von-Liebig-Weg 3, 37077 G\"ottingen, Germany \label{mps}
\and
National Solar Observatory, Boulder, CO 80303, USA \label{nso}
\and
Institut f\"ur Astrophysik und Geophysik, Georg-August-Universit\"at G\"ottingen,  37077 G\"ottingen, Germany \label{gottingen}
\and
Center for Astrophysics and Space Science, NYUAD Institute, New York University Abu Dhabi, Abu Dhabi, UAE \label{NYUAD}
\and
LIRA, Observatoire de Paris, Universit\'e PSL, Sorbonne Universit\'e, Universit\'e Paris Cit\'e, CY Cergy Paris Universit\'e, CNRS, 92190 Meudon, France
\label{ObsPM}
}

\date{Received $\langle$date$\rangle$ / Accepted $\langle$date$\rangle$}

\abstract
{}
{
We studied the temporal evolution of solar inertial modes over the solar cycle using observations from GONG and SDO/HMI.
We focused our attention on the high-latitude mode with azimuthal wavenumber $m=1$ and the equatorial Rossby modes with $3 \le m \le 16$.
}
{
We used maps of horizontal flows near the solar surface from the GONG and HMI (p-mode) ring-diagram pipelines at a cadence of approximately one day, covering the period 2002--2024.
The data were divided into overlapping 4-year windows, with central times separated by six months.
Within each time window and for each inertial mode, we measured the frequency and the power of the mode from the GONG and HMI data.
}
{
We find good agreement between the GONG and HMI measurements throughout their overlapping period from 2010 to 2024.
In general, the magnitude of the frequency variations increases with increasing $m$, and relative changes in mode power typically exceed 100\%.
For the $m=1$ high-latitude mode, the measured power is anticorrelated with the sunspot number, while its frequency shows no significant temporal variation.
For the equatorial Rossby modes, the frequencies are generally anticorrelated with the sunspot number, whereas the mode powers tend to correlate positively with the sunspot number. An exception is the $m=3$ equatorial Rossby mode, whose mode power is strongly anticorrelated with the sunspot number, in contrast to the other equatorial Rossby modes, highlighting its distinct behavior.
}
{
We find that the frequencies and power of the Sun's inertial modes have varied significantly on solar-cycle timescales over the past 23 years. The mode parameters are, however, not uniformly synchronized with the sunspot number; variations are observed both between modes and from one solar cycle to the next, with some of the differences likely reflecting observational uncertainties. The sensitivity of inertial modes to solar-cycle changes indicates their potential as a diagnostic of solar interior dynamics and magnetism.
}


\maketitle

\section{Introduction}

Equatorial Rossby (ER) modes on the Sun, also referred to as sectoral Rossby modes, were first identified observationally by \cite{Loeptien2018} for longitudinal wavenumbers $3 \leq m \leq 15$. Subsequent work by \cite{Gizon2021} and \citet{Hanson2022} revealed additional families of   inertial modes with maximum vorticity amplitudes peaking at mid and high latitudes. The high-latitude (HL) modes occur for $1 \leq m \leq 5$, with the $m=1$ mode dominating in amplitude ($\sim10$~m\,s$^{-1}$) and peaking above $60^\circ$ latitude. A recent overview of solar inertial modes is given by \cite{Gizon2024} and references therein.

The parameters of the Sun's inertial modes vary over the solar cycle. For the ER modes, \cite{Waidele2023} show that, in helioseismic flow maps from the Helioseismic and Magnetic Imager onboard the Solar Dynamics Observatory \citep[SDO/HMI:][]{Scherrer2012,Schou2012}, the average power of modes with $3 \leq m \leq 16$ is higher during the maximum of solar cycle 24 (around 2014) and lower during the solar minimum between cycles 24 and 25, while the average mode frequency decreases during solar maximum. In contrast, the $m=1$ HL mode exhibits an opposite amplitude trend: using direct Doppler observations over the last five solar cycles from HMI, the Global Oscillation Network Group \citep[GONG:][]{Harvey1996}, and the Mount Wilson Observatory, \cite{Liang2025} found that the mode power is anticorrelated with the solar cycle, whereas the mode frequency shows only weak variations and no clear cycle dependence.

As noted by \cite{Gizon2021} and \cite{Bekki2022linear}, numerical models indicate that solar inertial modes are sensitive to conditions throughout much of the convection zone. The solar-cycle variations observed in their frequencies are therefore likely to be caused by a combination of perturbations in the solar interior. The most prominent candidates are changes in the solar differential rotation \citep{Goddard2020} and variations in the global solar magnetic field. The stochastic excitation of the modes \citep[][]{Philidet2023} may also be affected indirectly  by the magnetic field.
By isolating and removing the effects associated with the well-understood changes in the differential rotation \citep[e.g.,][]{Vorontsov2002, Howe2009},  measurements of the solar-cycle variations of the inertial modes offer the exciting prospect of placing meaningful constraints on the Sun's global magnetic field in the bulk of the convection zone. This provides a strong motivation for detailed studies of inertial-mode variability.

In this paper we further characterize the parameters of solar inertial modes using ring-diagram (RD) flow maps spanning the GONG (2002--2024) and HMI (2010--2024) observing periods. By comparing these two independent datasets, we aim to confirm the reported frequency variations. In addition, we present time-dependent parameters for individual modes as well as averages over mode sets.

\section{Observations}
\label{sec:data}

\begin{figure*}[t]
    \centering
    \includegraphics[width=0.95\textwidth]{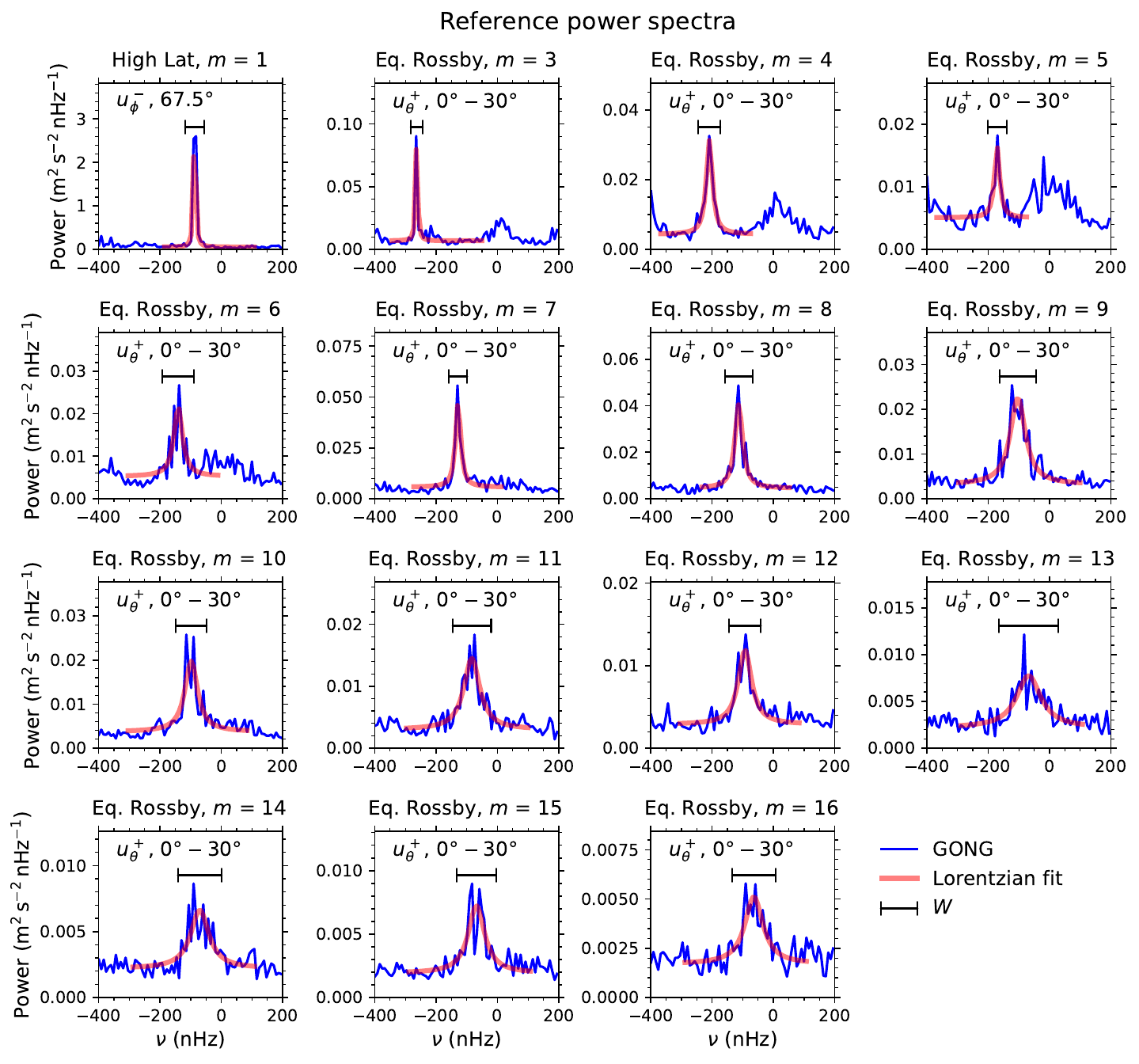}
    \caption{GONG reference power spectra $\overline{P}_m(\nu)$ (blue curves) with  frequency resolution  $\textrm{d}\nu=8$~nHz. The top-left panel shows   $\overline{P}_m(\nu)$ for the $m=1$ component of $u_{\phi}^-$; the peak corresponds to the  $m = 1$ HL mode.
    All other panels
    show   $\overline{P}_m(\nu)$ for the $u_{\theta}^+$ component of the flow and different $m$ values ranging from $3$ to $16$.
    The red curves show the Lorentzian fits to the power spectra (Sect.~\ref{sec:mean_spectra}). The horizontal black segments show the frequency windows ($W$) used to extract the mode parameters from consecutive 4-year power spectra (see Sect.~\ref{sec:parameter_calc}).}
    \label{fig:mean_power_spectra}
\end{figure*}

\subsection{Overview of the datasets}

We utilized near-surface zonal and meridional flow measurements derived from the RD pipelines of GONG \citep{Hill2003, Corbard2003} and HMI \citep{Bogart2011a, Bogart2011b}, spanning from January 2002 to September 2024 and July 2010 to September 2024, respectively. The RD pipeline provides flow measurements by analyzing shifts in p-mode frequencies obtained from tracked patches of Doppler velocity measurements \citep{Haber2002}.

In the GONG RD pipeline, full-disk Dopplergrams from multiple GONG sites are merged and remapped into overlapping patches of size $15^{\circ}\times15^{\circ}$, with centers separated by $7.5^{\circ}$. These patches cover latitudes and central meridian distances (CMDs) within $\pm 52.5^{\circ}$. For this study, flows at four additional latitudes, $\pm60^{\circ}$ and $\pm 67.5^{\circ}$, are also included. Each patch is tracked for 1664 minutes at the Snodgrass rotation rate \citep{Snodgrass1984}.

The HMI RD pipeline uses overlapping patches of the same size, tracked at the sidereal Carrington rotation rate. These patches cover latitudes and CMDs within $\pm 75^{\circ}$. The latitudinal spacing is identical to that of GONG, while the CMD separation is $7.5^{\circ}$ within $\pm30^{\circ}$ latitude and becomes coarser at higher latitudes. To ensure consistency, we interpolated the HMI flows in the CMDs \citep{Proxauf2020} to match the GONG grid points.
Accordingly, both GONG and HMI datasets are provided on a grid that covers latitudes from $-67.5^\circ$ to $67.5^\circ$ every $7.5^\circ$. The HL-mode analysis is restricted to the latitude bins centered at $\pm 67.5^\circ$ where the mode power is strongest, while the ER-mode analysis is restricted to latitude bins within $\pm30^\circ$.

\subsection{Post-processing of ring-diagram data}

The temporal cadence of GONG and HMI flow maps is, on average, about 27.2753 hours. To remove the annual and zero-frequency components from the GONG and  HMI measurements, we fit the time series with sinusoids and subtract the resulting fits from the flow maps in the Stonyhurst frame \citep{Proxauf2020}.

Since the GONG RD pipeline utilizes ground-based daily observations from different sites, an additional one-day periodicity is introduced into the data. The frequency corresponding to this one-day period is $11.6$~$\mu$Hz. Due to aliasing, this frequency appears at $1.4$~$\mu$Hz, as the Nyquist frequency for our dataset is $5.1$~$\mu$Hz. This aliased frequency corresponds to a periodicity of approximately $8.33$ days. We also removed this additional periodicity from the GONG data.

The zonal and meridional flows in the Stonyhurst frame are transformed into the Carrington frame to obtain $u_\phi(\theta,\phi,t)$ in the longitudinal direction and $u_\theta(\theta,\phi,t)$ in the co-latitudinal direction at the observation time $t$. The Carrington longitude ($\phi$) increases in the prograde direction, and the co-latitude ($\theta$) increases southward.

\section{Measurements of mode parameters}

\subsection{Mode parameters from mean spectra}
\label{sec:mean_spectra}

We divided the observation periods of GONG and HMI into overlapping time segments of size $T = 4$~years, with the centers shifted by multiples of six months. The central times of the GONG segments are given by $t_n = \text{1 January 2004} + n \times 6$ months, where $n$ ranges from 0 to 38, covering the period from 2002 to 2024. For HMI, the central times are $t_n = \text{1 July 2012} + n \times 6$ months, with $n$ ranging from 0 to 21, covering the period from mid 2010 to 2024.

The flow maps within each time segment are symmetrized ($+$) or anti-symmetrized ($-$) with respect to the equator, to obtain $u_j^{s}(\theta, \phi, t)$, where $s$ represents the symmetry and $j$ represents either $\theta$ or $\phi$. The ER modes are stronger in the $u_\theta^+$ component, while the $m=1$ HL mode is stronger in $u_\phi^{-}$. Therefore, we utilized these components to compute their respective power spectral density:
\begin{equation}
    P_j^s\,(\theta,m,\nu,t_n) = \eta_1 \eta_2 \left| \sum_{\phi,t^{\prime}} \textrm{rect}\left( \frac{t'-t_n}{T} \right)  u_{j}^{s}(\theta,\phi,t^{\prime})e^{\textrm{i}(m\phi-2\pi\nu t^{\prime})}\right|^2 ,
\label{eq:FFT}
\end{equation}
where the function ``rect'' refers to a rectangular window, defined as rect$(\xi)=1$ for $|\xi| < 1/2$ and 0 otherwise.
The azimuthal wavenumber $m$ is taken to be a positive integer throughout this paper.
The temporal frequency $\nu$ is measured in the Carrington frame (retrograde propagation occurs at $\nu < 0$).
The factor $\eta_1=N_\phi^2 N_t^2/({N_\phi'^2 N_t'^2})$ corrects for the reduction in power due to missing data in time and partial coverage in longitude. Here, $N_{\phi}$ and $N_{t}$ represent the total number of grid points in longitude and time within each time segment, while $N'_{\phi}$ and $N'_{t}$ denote the number of available data points. The factor  $\eta_2 = 2 / (\textrm{d}\nu\ N_\phi^2 N_t^2)$ is a normalization factor, where $\textrm{d}\nu = 8$~nHz is the frequency resolution. With this definition, the power $P_m$ has units of $\rm m^2 s^{-2} \, nHz^{-1}$.

To study the ER modes, we averaged $P_\theta^+$ over latitude bins within $\pm30^\circ$ of the equator, while for the HL mode, we used $P_\phi^-$ only at the highest latitude bin $67.5^\circ$. That is,
\begin{equation}
P_m(\nu,t_n) = \begin{cases}
    \;P_\phi^-(\theta=22.5^\circ,m,\nu,t_n) & \text{for HL mode,} \\
    \;\left\langle P_\theta^+(\theta,m,\nu,t_n) \right\rangle_{|90^\circ-\theta| \leq 30^\circ} & \text{for ER modes.} \rule{0pt}{3ex}
\end{cases}
\end{equation}
For each $m$, we computed a GONG reference power spectrum for the full time series covering 2002--2024, denoted by $\overline{P}_m(\nu)$. The frequency resolution of the reference power spectrum was binned down to 8~nHz, identical to that of the other 4-year power spectra analyzed in this paper.
The power spectrum associated with each inertial mode  is fitted with a Lorentzian profile,
\begin{equation}
    L_m(\nu)=
    \overline{A}_m \left[ 1 + \left(\frac{\nu-\overline{\nu}_m}{\overline{\Gamma}_m/2}\right)^{2} \right]^{-1} + \overline{B}_m ,
    \label{eq:Lorentzian}
\end{equation}
where $\overline{A}_m$, $\overline{\nu}_m$, $\overline{\Gamma}_m$, and $\overline{B}_m$ are the Lorentzian height, central frequency, full width at half maximum, and a constant background, respectively.
The mode parameters are estimated from the data using the maximum likelihood method (see Appendix~\ref{sec:Likelihood_function}).
For each mode, we chose a fitting range that excludes spatial leakage from neighboring $m$'s, as well as low-frequency power from active-region flows.
In their Table~1, \cite{Liang2019} provide the fitting ranges for the Rossby modes. For modes with $m \le 8$, there is little room for adjustment, given the need to avoid both spatial leakage and contamination by low-frequency power from inflows into active regions. We therefore adopted essentially the same fitting range as \citet{Liang2019} for these modes. For modes with $m>8,$ we extended the width of the fitting window from \citet{Liang2019}'s 300~nHz to 400~nHz to improve the estimate of the background. The fitted parameters depend  on the choice of fitting windows; however, we verified that  modest adjustments  do not affect the main results.
Table~\ref{tab:window} lists the fitting range for each mode.
We caution that mode frequencies
derived from Lorentzian fits may be slightly biased if the power spectrum associated with an inertial mode exhibits asymmetry about the peak frequency.

\begin{table}
\caption{Mode parameters measured from the GONG reference power spectra (2002--2024).}
\centering
\renewcommand{\arraystretch}{1.5}
\begin{tabular}{c c c c c c}
\hline \hline
 $m$ & $(j,s)$        & $\overline{\nu}_m$     & $\overline{\Gamma}_m$   & $\overline{E}_m$       & S/N \\ [-4pt]
     &                & (nHz)                  & (nHz)                   & (m$^2$~s$^{-2}$)       & \\
\hline
\multicolumn{6}{c}{\it{High-latitude mode}} \\
\hline
 $1$ & ($\phi$,$-$)   & $ -87.0_{-0.8}^{+0.9}$ & $ 4.8_{ -1.4}^{ +2.2}$ & $48.2_{-13.89}^{+14.24}$ & $118.0_{-57.8}^{+90.0}$ \\
\hline
\multicolumn{6}{c}{\it{Equatorial Rossby modes}} \\
\hline
 $3$ & ($\theta$,$+$) & $-262.9_{-0.8}^{+0.7}$ & $ 6.6_{- 1.6}^{+ 1.9}$ & $ 1.0_{-0.17}^{+0.17}$ & $13.7_{-4.1}^{+5.7}$ \\
 $4$ & ($\theta$,$+$) & $-208.8_{-1.7}^{+1.7}$ & $24.0_{- 4.5}^{+ 5.0}$ & $ 1.0_{-0.12}^{+0.13}$ & $ 6.2_{-1.1}^{+1.5}$ \\
 $5$ & ($\theta$,$+$) & $-170.7_{-2.5}^{+2.5}$ & $20.3_{- 6.7}^{+ 7.0}$ & $ 0.4_{-0.08}^{+0.09}$ & $ 2.2_{-0.5}^{+0.9}$ \\
 $6$ & ($\theta$,$+$) & $-140.5_{-2.8}^{+2.7}$ & $34.0_{- 7.8}^{+ 8.6}$ & $ 0.9_{-0.13}^{+0.15}$ & $ 3.1_{-0.5}^{+0.8}$ \\
 $7$ & ($\theta$,$+$) & $-127.9_{-1.5}^{+1.5}$ & $19.9_{- 3.7}^{+ 3.9}$ & $ 1.3_{-0.16}^{+0.17}$ & $ 7.3_{-1.4}^{+1.9}$ \\
 $8$ & ($\theta$,$+$) & $-113.0_{-1.8}^{+1.8}$ & $29.9_{- 4.8}^{+ 5.2}$ & $ 1.7_{-0.17}^{+0.19}$ & $ 8.1_{-1.2}^{+1.7}$ \\
 $9$ & ($\theta$,$+$) & $-103.0_{-2.7}^{+2.7}$ & $59.7_{- 8.4}^{+ 9.0}$ & $ 1.8_{-0.16}^{+0.19}$ & $ 6.1_{-0.8}^{+1.1}$ \\
$10$ & ($\theta$,$+$) & $ -98.1_{-2.9}^{+2.9}$ & $50.5_{- 8.4}^{+ 9.0}$ & $ 1.3_{-0.14}^{+0.16}$ & $ 4.3_{-0.6}^{+0.8}$ \\
$11$ & ($\theta$,$+$) & $ -83.3_{-3.3}^{+3.3}$ & $62.9_{-10.6}^{+11.0}$ & $ 1.1_{-0.13}^{+0.15}$ & $ 3.9_{-0.5}^{+0.7}$ \\
$12$ & ($\theta$,$+$) & $ -92.7_{-3.3}^{+3.4}$ & $52.2_{-10.0}^{+10.4}$ & $ 0.7_{-0.09}^{+0.11}$ & $ 3.2_{-0.4}^{+0.7}$ \\
$13$ & ($\theta$,$+$) & $ -67.8_{-5.0}^{+5.0}$ & $96.0_{-18.8}^{+20.7}$ & $ 0.8_{-0.13}^{+0.16}$ & $ 2.6_{-0.3}^{+0.6}$ \\
$14$ & ($\theta$,$+$) & $ -70.3_{-4.8}^{+4.8}$ & $71.8_{-16.3}^{+17.9}$ & $ 0.5_{-0.08}^{+0.10}$ & $ 2.0_{-0.3}^{+0.5}$ \\
$15$ & ($\theta$,$+$) & $ -68.3_{-3.9}^{+3.9}$ & $64.6_{-12.5}^{+13.5}$ & $ 0.5_{-0.07}^{+0.09}$ & $ 2.8_{-0.4}^{+0.6}$ \\
$16$ & ($\theta$,$+$) & $ -63.7_{-4.9}^{+4.8}$ & $71.3_{-16.3}^{+17.8}$ & $ 0.4_{-0.06}^{+0.08}$ & $ 2.0_{-0.3}^{+0.5}$ \\
\hline
\end{tabular}
\tablefoot{Frequencies ($\overline{\nu}_m$), linewidths ($\overline{\Gamma}_m$), mode powers ($\overline{E}_m = \pi\overline{A}_m\overline{\Gamma}_m/2$), and signal-to-noise ratios ($\text{S/N} = \overline{A}_m/\overline{B}_m$) for  each mode. The corresponding flow component and symmetry used to compute $\overline{P}_m(\nu)$ are denoted by $j=\theta, \phi$ and $s = +, -$, respectively. The upper and lower bounds indicate the $68\%$ confidence levels of the parameters obtained from Monte Carlo simulations.}
\label{tab:parameters}
\end{table}

\begin{figure}[t]
\centering
\resizebox{0.82\hsize}{!}{\includegraphics{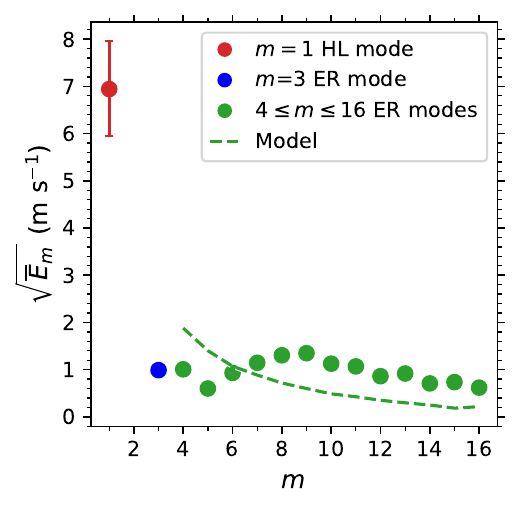}}
\caption{Square root of mode power as a function of $m$ for the GONG observations ($\overline{E}_m$ from Table~\ref{tab:parameters}). The stochastic excitation model from \cite{Philidet2023} is overplotted.}
\label{fig:power_model_obs}
\end{figure}

\begin{figure*}[h!]
    \centering
    \includegraphics[width=0.96\textwidth]{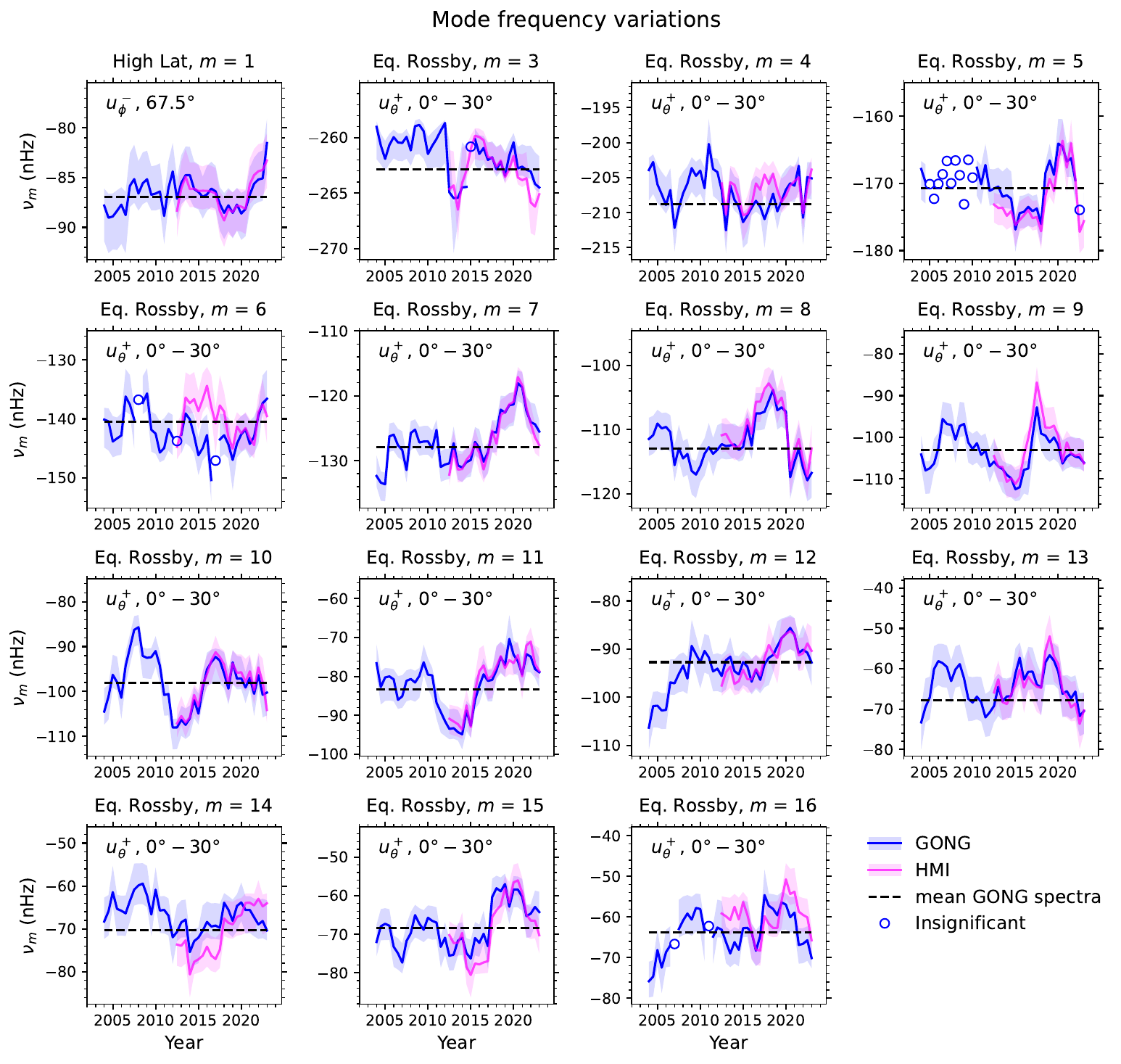}
    \caption{Temporal variations of the mode frequencies ($\nu_m$) obtained from the GONG (blue) and HMI (magenta) datasets.
    The top-left panel is for the $m=1$ HL mode, the other panels for the ER modes.
    The shaded regions indicate the $68\%$ confidence intervals of $\nu_m$ estimated from Monte Carlo simulations.   Open circles indicate mode frequencies measured in time segments with no significant power ($<95\%$ confidence).  The mode frequencies obtained from the reference GONG data (2002--2024) are shown as horizontal dashed lines.}
    \label{fig:mode_frequency}
\end{figure*}

\begin{figure*}[h]
    \centering
    \includegraphics[width=0.96\textwidth]{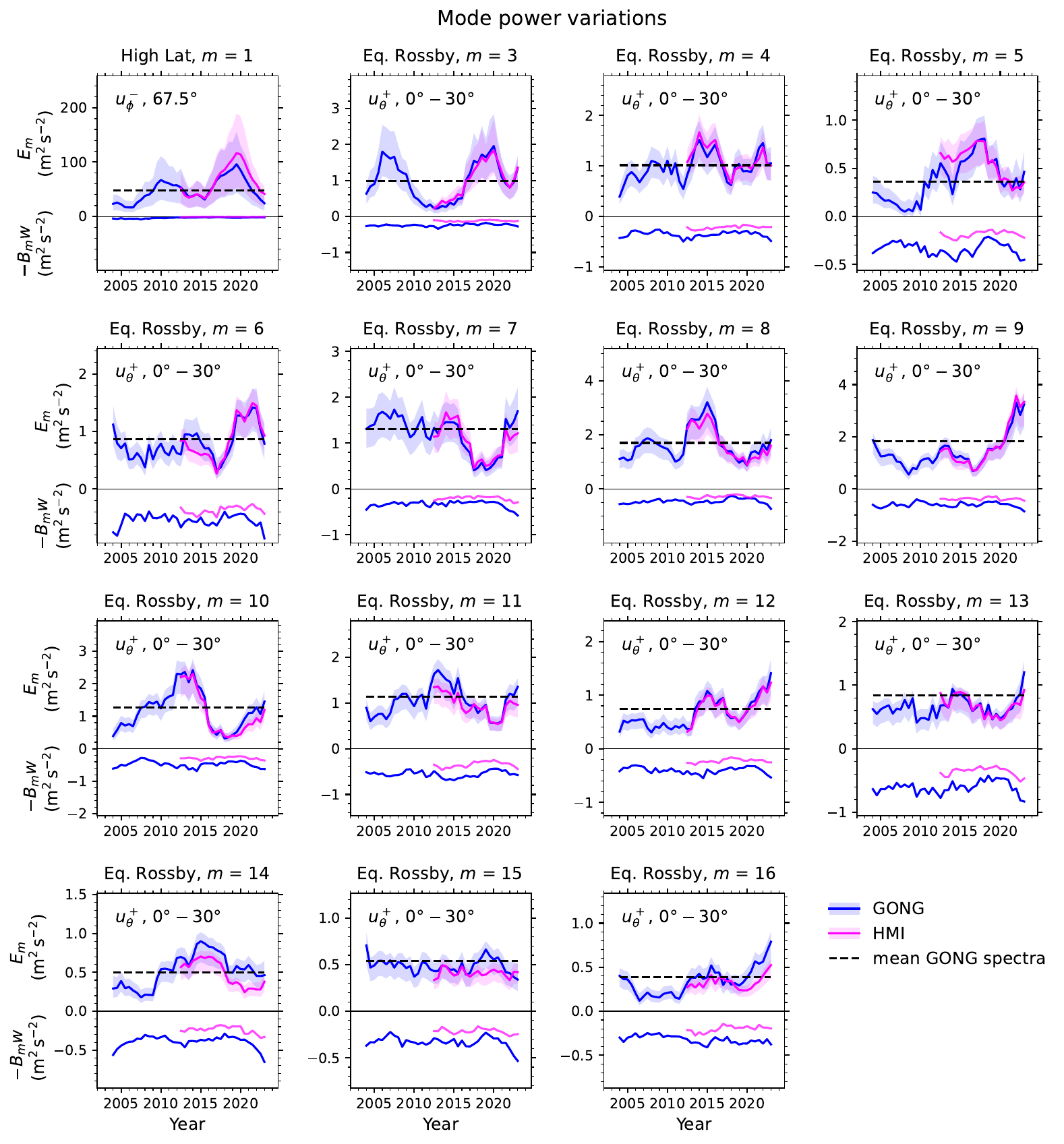}
    \caption{Temporal variations of the mode power ($E_m$) and the negative of the background power ($-B_m w$). The top-left panel is for the $m=1$ HL mode, the other panels for the ER modes. The powers computed from the GONG and HMI datasets are shown in blue and magenta, respectively. The shaded regions represent the $68\%$ confidence intervals of $E_m$ estimated from Monte Carlo simulations. The horizontal dashed lines indicate the mode powers obtained from the reference GONG data (2002--2024).
    }
    \label{fig:mode_power}
\end{figure*}

We defined the mode power as the integrated  power under the Lorentzian profile (minus the background power):
\begin{equation}
    \overline{E}_m = \frac{\pi}{2} \, \overline{A}_m\,\overline{\Gamma}_m.
    \label{eq.power}
\end{equation}
Table~\ref{tab:parameters} provides the mode parameters and Fig.~\ref{fig:mean_power_spectra} shows the GONG reference spectra, $\overline{P}_m(\nu)$, along with the Lorentzian fits for $m=1$ HL and $3 \leq m \leq 16$ ER modes. A comparison between the mean GONG and HMI spectra for the period 2010.5--2022.5 is shown in Fig.~\ref{fig:meanspectra_HMI_GONG}, demonstrating that the two datasets are consistent with each other.
For reference, the mode frequencies listed in Table~\ref{tab:parameters} are compared in Fig.~\ref{fig:cf-freq} with previous measurements reported by various authors. These differences arise primarily from the differing temporal coverage of the datasets.

The measured mean mode powers are plotted in
Fig.~\ref{fig:power_model_obs} for all modes of interest. The ER modes have powers in the range $0.4$--$1.8$~m$^2$~s$^{-2}$.
A comparison is provided in
Fig.~\ref{fig:power_model_obs} between the measured mode powers and the model of \citet{Philidet2023}; this 2D model in the $\beta$-plane is based on the stochastic excitation of purely toroidal modes by turbulent convection. The agreement between observations and model is reasonable.
The power of the $m=1$ HL mode stands out at $48.2$~m$^2$~s$^{-2}$ in the average GONG spectrum. This value may be related  to the  HMI maximum velocity amplitude of $9.8$~m~s$^{-1}$ above $67.5^\circ$ reported by \citet{Gizon2021}.
The mode power varies strongly with time. The temporal average of the mode power depends strongly on the start and end of each dataset. For example, the $m=1$ HL mode had much lower power in 2002--2008 than during quiet-Sun periods. Therefore, the average mode power that we find for the period 2002--2024 is lower than the value reported by \citet{Gizon2021} for the period 2010--2020.

The solar $m = 1$ HL mode is self-excited by a baroclinic instability \citep{Bekki2024}, thus explaining why it reaches such a high amplitude in the Sun. However, all the other modes discussed here are linearly stable, and their excitation is due instead to the highly turbulent motions characterizing the solar convection zone, in a manner similar to the stochastic excitation of the solar p-modes \citep{Philidet2023}.

\subsection{Mode parameters from individual time segments}
\label{sec:parameter_calc}

We measured the mode frequencies and powers in individual 4-year time segments to study their temporal variations.
Because some modes are not significant in certain periods, the Lorentzian fit may not perform well.
We therefore developed a more robust approach for determining the mode parameters.

We defined the  mode frequency ($\nu_m)$ and the mode power ($E_m$) in each time segment ($t_n$) as
\begin{align}
    \nu_m(t_n) &= \frac{\sum_{\nu\in W} \nu \times P_m(\nu,t_n) }{\sum_{\nu\in W} P_m(\nu,t_n)} ,
    \label{eq:frequency_calc} \\
    E_m (t_n) &=
    \frac{(\pi/2)\textrm{d}\nu}{\arctan(w/\overline{\Gamma}_m)}\sum_{\nu\in W} \left[ P_m(\nu,t_n) - B_m(t_n) \right],
    \label{eq:power_calc}
\end{align}
where $W$ is the frequency window over which the sums are computed (see Table~\ref{tab:window}, right column), and $w$ is the width of the frequency window.
The window $W$, centered at the mean frequency $\overline{\nu}_m$ and with a width of at least $2\overline{\Gamma}_m$, is selected to encompass the majority of the mode power.
The normalization factor $\arctan(w/\overline{\Gamma}_m)/(\pi/2)$ is introduced so that $E_m$ can be directly compared to the product $\pi A_m\Gamma_m/2$ (correction for the missing power in the wings of the Lorentzian beyond the interval $W$).
The background power $B_m(t_n)$ is estimated as the median power away from the mode (outside the interval $W$) and within the fitting range given in Table~\ref{tab:window}.

\begin{figure*}
    \centering
    \includegraphics[width=0.99\textwidth]{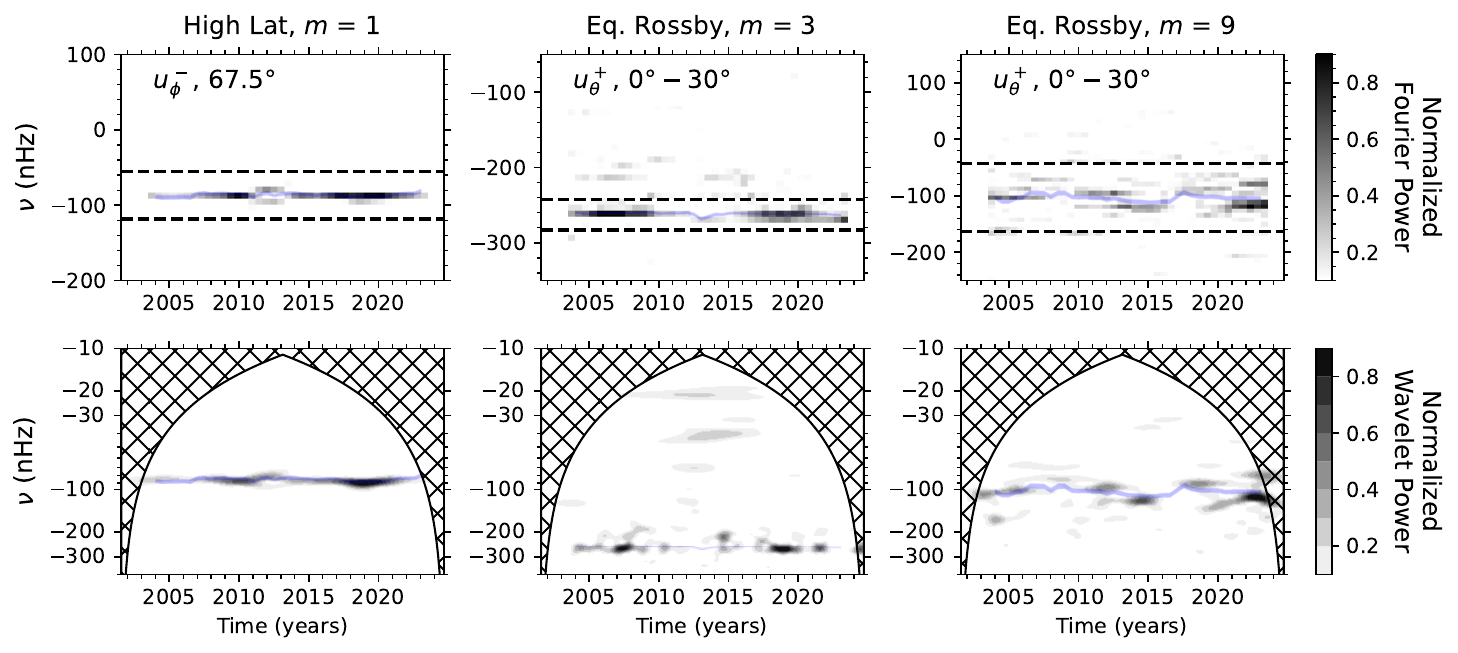}
    \caption{Fourier (top panels) and wavelet (bottom panels) power spectra of the $m=1$ HL and $m=3$ and $9$ ER modes, as a function of time and frequency. The frequencies $\nu_m(t_n)$, calculated using Eq.~\eqref{eq:frequency_calc}, are overplotted as light blue curves, with widths representing the $68\%$ confidence intervals. In the top panels, the horizontal dashed lines indicate the frequency window ($W$) used in Eqs.~\eqref{eq:frequency_calc} and \eqref{eq:power_calc}.}
\label{fig:2D_spectra}
\end{figure*}

The values of $\nu_m$ and $E_m$ are shown in Figs.~\ref{fig:mode_frequency} and~\ref{fig:mode_power}. The parameters obtained from GONG and HMI are consistent during their overlap period (2010--2024). Furthermore, the mode powers and frequencies exhibit significant temporal variations in most cases.
Figure~\ref{fig:mode_power} also shows the background power variations. It is evident that the mode power variations are not caused by changes in the background power. The background power of GONG is higher than that of HMI, likely due to a larger number of missing data points.

We estimated the uncertainties in $\nu_m$ and $E_m$ using Monte Carlo simulations. We generated 10\,000 realizations of a model spectrum with linewidth $\Gamma_m = \overline{\Gamma}_m$, amplitude $A_m = E_m/(\pi\,\overline{\Gamma}_m/2)$, frequency $\nu_m$, and background $B_m$. The $68\%$ confidence intervals of the resulting parameter distributions are taken as the uncertainty bounds for the $\nu_m$ and $E_m$.

For a subset of modes, Lorentzian fits performed on 4-year time segments are robust. These include the $m=1$ HL mode, and the $m=4$, $8$, $12$, and $14$ ER modes. For these five modes, we were able to extract reliable temporal variations of the mode line widths ($\Gamma_m$) in the sense that the product of the fitted Lorentzian amplitude and line width is consistent with the independently measured mode power ($E_m$) obtained using the method described above. The results are shown in Appendix~\ref{fig:lorentzianfit}.
For the remaining modes, the Lorentzian fits are not sufficiently stable across the 4-year segments, and the resulting parameters do not show satisfactory consistency with the corresponding mode power ($E_m$). We therefore do not consider these fits reliable and do not report these modes from the analysis based on Lorentzian fitting.

\begin{figure*}
\sidecaption
    \includegraphics[width=11cm]{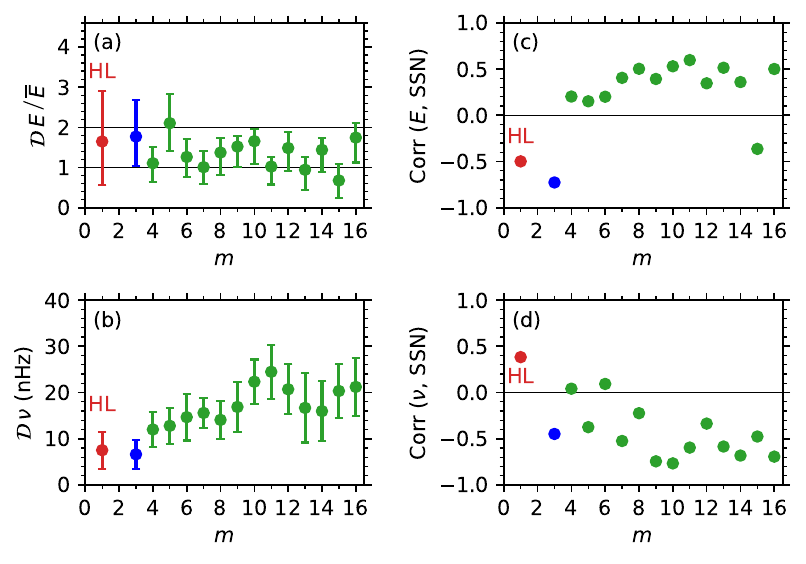}
    \caption{Solar cycle variability of mode parameters.
    (a) Relative peak-to-peak variation of mode powers over the GONG period 2002--2024.
    Error bars indicate the 68\% confidence intervals.
    (b) Peak-to-peak variation of mode frequencies over the GONG period 2002--2024.
    (c) Correlation coefficients between the time series of mode powers and the SSNs, computed over the full GONG period.
    (d) Same as panel (c) but for mode frequencies.
    The values for the $m=1$ HL  mode are shown in red, those for the $m\ge 4$ ER modes in green. The $m=3$ ER mode is shown in blue.
    \vspace{0.3cm}
    }
    \label{fig:power_freq_ssn}
\end{figure*}

\subsection{Comparison of power spectra from Fourier and wavelet analyses}

The top panels of Fig.~\ref{fig:2D_spectra} show the latitudinally averaged Fourier power, $P_m(\nu,t_n)$, for each 4-year time segment of the $m=1$ HL mode and the $m=3$ and $m=9$ ER modes, as a function of time and frequency.
The frequencies $\nu_m(t_n)$, calculated using Eq.~\eqref{eq:frequency_calc}, are plotted over the Fourier spectra for comparison. The measured $\nu_m$ closely follow the excess power within the frequency window $W$.
The time periods during which the excess power peaks agree well with those shown in Fig.~\ref{fig:mode_power}.

We also performed a continuous wavelet transform using a Morlet wavelet. The bottom panels of Fig.~\ref{fig:2D_spectra} show the wavelet power of the selected modes within the inertial frequency ranges. The wavelet spectra are consistent with the Fourier spectra, though they appear more smeared.
The Fourier and wavelet spectra for all remaining modes are shown in Fig.~\ref{fig:2D_spectra_allmodes}.

\section{Temporal variations of mode parameters}

\subsection{Peak-to-peak variations of mode parameters}

We first examined the peak-to-peak variations of the mode parameters across the full GONG time series (2002--2024). The peak-to-peak variation in mode power is defined as
\begin{equation}
    \mathcal{D} E_m \equiv \max_n{[E_m(t_n)]} - \min_n{[E_m(t_n)]}.
\end{equation}
Analogously, the peak-to-peak variation of the frequency of a mode $m$ is denoted by $\mathcal{D} \nu_m$.

The values of $\mathcal{D} E_m$ are typically between 100\% and 200\%, as can be seen in Fig.~\ref{fig:power_freq_ssn}a. Most modes fall within this range, with a few modes showing values slightly below 100\%.
Although the $m=1$ HL mode has a power that is two orders of magnitude larger than ER modes (see Table~\ref{tab:parameters}), its relative peak-to-peak variation is of the same order as that of the ER modes.

Figure~\ref{fig:power_freq_ssn}b shows the peak-to-peak variations ($\mathcal{D}\nu_m$).  The least significant $\mathcal{D}\nu_m$ is for the $m=1$ HL mode. In all other cases $\mathcal{D}\nu_m$ is above three times its associated error.
The values of $\mathcal{D}\nu_m$ increase with increasing $m$ up to $m\approx 10$; for $m\gtrsim 10$, there seems to be a plateau around $\mathcal{D}\nu_m \approx 20$~nHz.
For low $m$ values, the peak-to-peak variations in $\nu$ correspond to less than about 10\% of their mean frequencies in the corotating frame, which is much smaller than the relative mode power variations.

\subsection{Correlations with sunspot number}

We calculated the Pearson correlation coefficients between the time series of mode parameters and the sunspot number (SSN).
The SSN data, obtained from the World Data Center SILSO, Royal Observatory of Belgium \citep{SILSO_Sunspot_Number}, were averaged over 4-year time segments for comparison with our data.
The correlation coefficients are listed in Table~\ref{tab:corr} for the full GONG period (2002--2024) and for the HMI-GONG overlap period (2010--2024).

\begin{table}
\caption{Pearson correlation coefficients between SSN and mode parameters.}
\centering
\renewcommand{\arraystretch}{1.2}
\begin{tabular} {r r r c r r}
\hline \hline
     & \multicolumn{2}{c}{GONG 2002--2024} & & \multicolumn{2}{c}{GONG 2010--2024} \\
\cmidrule(lr){2-3} \cmidrule(lr){5-6}
     & \multicolumn{2}{c}{Correlations} & & \multicolumn{2}{c}{Correlations} \\
 $m$ & ($E$, SSN)\!\!\! & ($\nu$, SSN)\!\!\! & & ($E$, SSN)\!\!\! & ($\nu$, SSN)\!\!\! \\
\hline
\multicolumn{6}{c}{\it High-latitude mode} \\
\hline
 $1$ & $-0.50$ & $ 0.38$ & & $-0.91$ & $ 0.78$ \\
\hline
\multicolumn{6}{c}{\it Equatorial Rossby modes} \\
\hline
 $3$ & $-0.73$ & $-0.45$ & & $-0.86$ & $-0.64$ \\
 $4$ & $ 0.20$ & $ 0.04$ & & $ 0.55$ & $-0.03$ \\
 $5$ & $ 0.15$ & $-0.37$ & & $-0.34$ & $-0.46$ \\
 $6$ & $ 0.20$ & $ 0.09$ & & $-0.03$ & $ 0.47$ \\
 $7$ & $ 0.41$ & $-0.52$ & & $ 0.91$ & $-0.69$ \\
 $8$ & $ 0.50$ & $-0.22$ & & $ 0.81$ & $-0.60$ \\
 $9$ & $ 0.39$ & $-0.74$ & & $ 0.23$ & $-0.70$ \\
$10$ & $ 0.53$ & $-0.77$ & & $ 0.86$ & $-0.71$ \\
$11$ & $ 0.60$ & $-0.60$ & & $ 0.90$ & $-0.76$ \\
$12$ & $ 0.35$ & $-0.34$ & & $ 0.33$ & $-0.53$ \\
$13$ & $ 0.51$ & $-0.58$ & & $ 0.74$ & $-0.59$ \\
$14$ & $ 0.36$ & $-0.68$ & & $ 0.22$ & $-0.72$ \\
$15$ & $-0.36$ & $-0.48$ & & $-0.72$ & $-0.75$ \\
$16$ & $ 0.50$ & $-0.69$ & & $ 0.41$ & $-0.80$ \\
\hline
\end{tabular}
\tablefoot{Correlation coefficients are computed from the GONG dataset for the full GONG period (2002--2024) and for the GONG period that overlaps with HMI (2010--2024).}
\label{tab:corr}
\end{table}

Figure~\ref{fig:power_freq_ssn}c shows the correlation coefficients between the mode powers and the SSN over the full GONG period for each $m$.
The mode power of the $m = 1$ HL mode has an anticorrelation of $-0.5$ with the SSN. We note that \citet{Liang2025} also report a correlation of $-0.5$ based on observations spanning five solar cycles.
The $m=3$ ER mode stands out in Fig.~\ref{fig:power_freq_ssn}c, exhibiting the strongest anticorrelation of $-0.7$, in contrast to the positive correlation coefficients of most other ER modes. We therefore highlight the $m=3$ ER mode with a different color in this plot.

Figure~\ref{fig:power_freq_ssn}d shows the correlation coefficients between the mode frequencies and the SSN.
In general, the ER-mode frequencies are anticorrelated with the SSN. For $m \gtrsim 9$, the correlations mostly fall between $-0.4$ and $-0.6$, whereas at smaller $m$ they are more scattered, with values between $0.1$ and $-0.5$.
The $m=3$ mode shows a negative correlation of $-0.4$  with the SSN. Although the $m=1$ HL mode shows a positive correlation of $0.4$, the frequency variations are almost always  within the uncertainties (see the top-left panel of Fig.~\ref{fig:mode_frequency}) and should not be overinterpreted.

\begin{figure}
\center
  \resizebox{0.95\hsize}{!}{\includegraphics{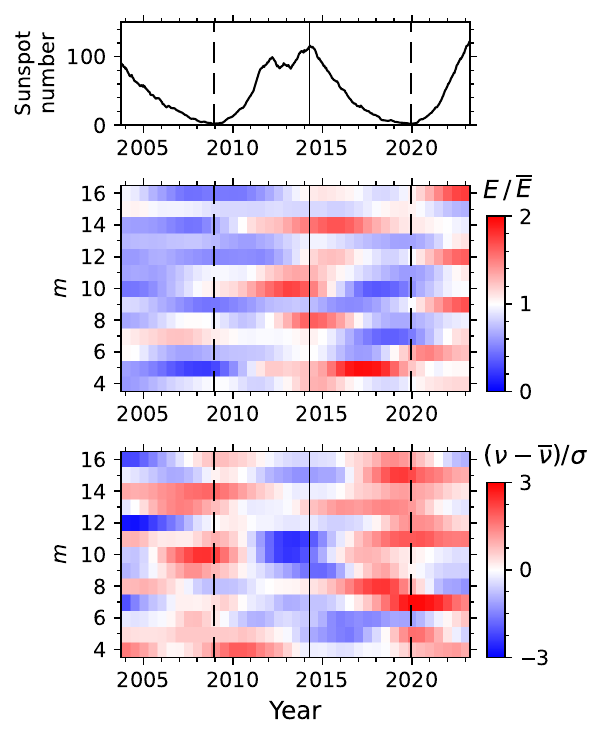}}
  \caption{
  Normalized ER mode powers (middle) and frequency shifts in units of frequency uncertainties (bottom) for $4 \le m \le 16$.
  The top panel shows the SSN in the corresponding time period.
  The GONG data are smoothed using a 3-year boxcar for clarity.
  }
  \label{fig:m-vs-t}
\end{figure}

\subsection{Mode power variations}

Since the powers of most of the ER modes (except the $m=3$) show a positive correlation with the SSN over the full GONG period, we studied these modes together.
To investigate whether there are common patterns, we stacked the time series of the mode powers over $m$.
The middle panel of Fig.~\ref{fig:m-vs-t} shows the temporal variations in the normalized mode power, $E_m/\overline{E}_m$, for $4 \le m \le 16$ ER modes.
Although they are positively correlated with the SSN, they do not peak exactly at the solar maximum.
The modes with $m=10$ and $11$ peak around the solar maximum, slightly toward the rising phase, whereas modes with lower and higher $m$ tend to peak during the declining phase.

\begin{figure*}
\sidecaption
    \includegraphics[width=12cm]{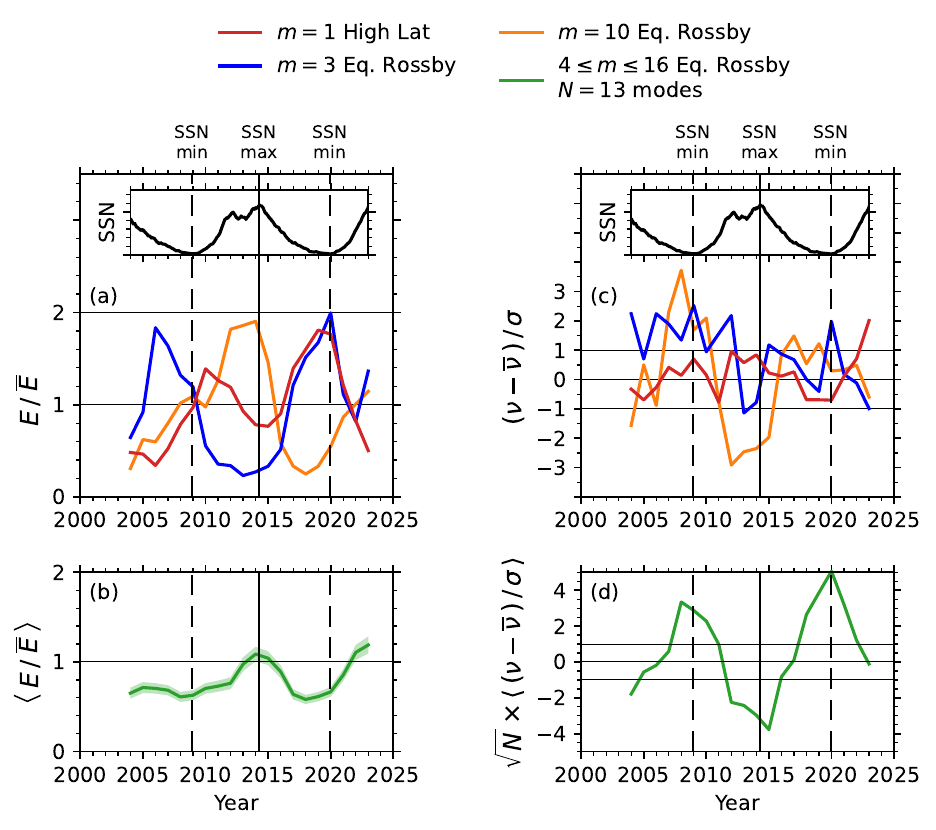}
    \caption{
    Temporal variations of solar inertial mode parameters.
    (a) Relative mode power for three selected modes.
    (b) Relative mode power averaged over ER modes with $4 \le m \le 16$; see Eq.~\eqref{eq:mean_power}.
    (c) Mode frequency shifts divided by their uncertainties for the three modes shown in panel (a).
    (d) Mode frequency shifts divided by their uncertainties averaged over $N=13$ ER modes with $4 \le m \le 16$, and multiplied by $\sqrt{N}$ to account for the noise reduction. In all panels, only every other data point is plotted for clarity.
    \vspace{-0.05cm}
    }
    \label{fig:summary}
\end{figure*}

Panels~a and~b of Fig.~\ref{fig:summary} compare the normalized power for the $m=1$ HL mode, the $m=3$ and $m=10$ ER modes, and the average over $4 \le m \le 16$ for the ER modes.
The averaged power is defined as
\begin{equation}
    \langle E/\overline{E} \rangle = \sum\limits_{m=4}^{16} \alpha_m(t)\    \frac{E_m(t)}  {\overline{E}_m }  ,
 \quad  \alpha_m := \frac{1/\sigma_{E_m}^2} { \sum_{m'} 1/\sigma_{E_{m'}}^2},
\label{eq:mean_power}
\end{equation}
where $\sigma_{E_m}$ is the uncertainty in $E_m$, normalized by ${\overline{E}_m}$.

Although the $m=1$ HL and $m=3$ ER mode powers are both anticorrelated with the SSN over the full GONG period, their $E_m/\overline{E}_{m}$ time series show patterns similar to those of the $4 \le m \le 16$ ER modes, in that the power does not peak exactly at the phase implied by the correlation.
Both modes peak near the cycle 24/25 minimum, but their peak power is offset from the cycle 23/24 minimum earlier in the time series.
This temporal offset weakens the overall anticorrelations compared with those obtained during the HMI-GONG overlap period, suggesting that the timing of peak power varies from cycle to cycle.

As a representative example of the $4 \le m \le 16$ range, the $m=10$ ER mode shows a temporal evolution similar to that of the ER-set, but its peak power occurs closer to the rising phase of the solar cycle.
Meanwhile, the averaged power, $\langle E/\overline{E} \rangle$, over $4 \le m \le 16$ shows a strong correlation with the SSN, with coefficients of $0.76$ for the full GONG period (2002--2024) and $0.89$ for the HMI-GONG overlap period (2010--2024).
The lower correlation over the longer time interval reflects the impact of cycle-to-cycle phase shifts.

During the solar activity cycle, strong magnetic fields may inhibit convection, thus affecting the efficiency with which turbulence in the convection zone drives the inertial modes, and directly impacting their excitation rate. At the same time, the turbulent viscosity, being tightly related to the convective turbulent spectrum, is also likely to have a solar-cycle dependence.
Because the Rossby mode power is the result of a balance between driving and damping processes (namely the ratio between its excitation rate and damping rate, as illustrated by Eq.~(\ref{eq.power})), its observed solar-cycle dependence is a valuable signature of how the convection, alongside the internal magnetic field and the torsional oscillation, evolves along the activity cycle. In this context, numerical simulations constitute a very useful tool for investigating the behavior of inertial modes, and in particular the degree to which they are excited under different physical conditions \citep{Bekki2022nonlinear, Blume2024, Fuentes2026}.

\subsection{Mode frequency variations}

The bottom panel of Fig.~\ref{fig:m-vs-t} shows the temporal variations of the mode frequency shifts, normalized by their respective uncertainties, $(\nu_m - \overline{\nu}_m)/\sigma_{\nu_m}$, for the $4 \le m \le 16$ ER modes, where $\sigma_{\nu_m}$ denotes the uncertainty in $\nu_m$.
Overall, the frequency shifts are anticorrelated with the SSN; however, their temporal evolution does not consistently coincide with solar minima, particularly during the latter half of the time series.
While several modes exhibit positive frequency shifts near the cycle~23/24 minimum, similar enhancements occur at different phases in the subsequent cycle, indicating that the timing of the extrema is not fixed from cycle to cycle.

Panels c and d of Fig.~\ref{fig:summary} compare the normalized frequency shifts $(\nu_m - \overline{\nu}_m)/\sigma_{\nu_m}$ for the $m = 1$ HL mode and the $m = 3$ and $m = 10$ ER modes, together with the frequency shift averaged over $4 \le m \le 16$.
The frequency variations of the $m=1$ HL and $m=3$ ER modes remain mostly within $\pm\sigma_{\nu_m}$ and therefore have a low signal-to-noise ratio over the full GONG period.
As a representative example within the set of  $4 \le m \le 16$ ER modes, the $m=10$ ER mode exhibits a clearer and more coherent temporal evolution than the low-$m$ modes.
Its first positive frequency shift coincides with the cycle~23/24 minimum, consistent with the overall anticorrelation with the SSN.
However, the second positive frequency shift does not occur near the subsequent minimum but instead appears during the declining phase of the solar cycle.
This behavior illustrates that even for ER modes with relatively high signal-to-noise ratio, the phase of the frequency extrema can vary from one cycle to the next.

In contrast, the frequency shift averaged over $4 \le m \le 16$ shows a very significant variation ranging from $-4$ to $5$ times the standard deviation of the reduced noise.
The average frequency shift is strongly anticorrelated with the SSN, with correlation coefficients of $-0.85$ over the full GONG period (2002--2024) and $-0.87$ over the HMI-GONG overlap period (2010--2024).
As in the case of the mode power, the reduced correlation over the longer time interval reflects cycle-to-cycle variations in the timing of the frequency extrema.

\section{Conclusion}
Using nearly 23 years of helioseismic flow measurements from GONG and HMI, we investigated the temporal variations in the power and frequency of the $m=1$ HL  mode and the ER  modes with $3 \leq m \leq 16$. Most modes exhibit significant temporal variability in their measured parameters, and the results obtained from the GONG and HMI datasets are consistent throughout their overlapping period (2010--2024).

For the GONG period 2002--2024, we find that the power of the $m=1$ HL mode is anticorrelated with the SSN, with a correlation coefficient of $-0.5$ \citep[consistent with][]{Liang2025}. In contrast, most ER modes exhibit a positive correlation with the SSN, in general agreement with earlier studies \citep{Waidele2023}. An exception is the $m=3$ ER mode, which shows a strong anticorrelation with the SSN (correlation coefficient $-0.7$), highlighting its distinctive behavior within the ER-mode family. This behavior is perhaps not unexpected, as the $m=3$ mode lies at the intersection of the HL and ER branches in the dispersion diagram \citep{Gizon2021} and may therefore possess a mixed character.

For the mode frequencies during the GONG period 2002--2024, individual ER modes with $m\gtrsim 10$ show larger variations ($\sim 20$~nHz, Fig.~\ref{fig:power_freq_ssn}b) that are anticorrelated with the SSN (correlation coefficients around  $-0.5$, Fig.~\ref{fig:power_freq_ssn}d), while modes with smaller $m$ show weaker correlations.
When the ER modes with $m\ge 4$ are considered together, the average frequency variation is highly significant at the $5\sigma$ level and is strongly anticorrelated with the SSN (correlation coefficient $-0.8$, Fig.~\ref{fig:summary}d).

We further find, based on the GONG data, that the variations in mode parameters differ from one solar cycle to the next (Fig.~\ref{fig:m-vs-t}). This behavior is also reflected in the correlations with the SSNs: the correlation coefficients are not the same when computed over the full GONG time span (23 years) as compared to the shorter HMI period (14 years); see Table~\ref{tab:corr}.

Overall, the results demonstrate that both the power and frequencies of solar inertial modes undergo substantial solar-cycle variations, with noticeable differences from mode to mode.
The differing correlations of mode power and frequency with the SSN, together with the lack of strict phase coherence among modes, suggest that individual modes have different sensitivities to the underlying physical processes -- differential rotation and magnetic fields.

\section*{Data availability}
The measurements  in this study are available online in the Edmond repository of the Max Planck Society at
\url{https://doi.org/10.17617/3.S5TXER}.

\begin{acknowledgements}
BL, Z-CL and LG designed research, BL and Z-CL measured the mode parameters, and all authors contributed to the final manuscript.
Data were acquired by GONG instruments operated by NISP/NSO/AURA/NSF with contribution from NOAA. The HMI data are courtesy of NASA/SDO and the HMI Science Team. We acknowledge partial support from ERC Synergy Grant WHOLESUN~810218. The data were processed at the German Data Center for SDO (GDC-SDO), funded by the German Aerospace Center (DLR). L.G. acknowledges support from the NYU Abu Dhabi Center for Astrophysics and Space Science.
Note: After acceptance of this paper, we became aware that the conclusion derived in Appendix~\ref{sec:Likelihood_function} had previously been obtained by Appourchaux (2003), A\&A, vol. 412, pp. 903--904.

\end{acknowledgements}

\bibliographystyle{aa}
\bibliography{references.bib}

@ARTICLE{Blume2024,
       author = {{Blume}, Catherine C. and {Hindman}, Bradley W. and {Matilsky}, Loren I.},
        title = "{Inertial Waves in a Nonlinear Simulation of the Sun's Convection Zone and Radiative Interior}",
      journal = {\apj},
         year = 2024,
        month = may,
       volume = {966},
       number = {1},
          eid = {29},
        pages = {29},
          doi = {10.3847/1538-4357/ad27d1},
archivePrefix = {arXiv},
       eprint = {2312.14270},
 primaryClass = {astro-ph.SR},
       adsurl = {https://ui.adsabs.harvard.edu/abs/2024ApJ...966...29B}
}

@ARTICLE{Fuentes2026,
       author = {{Fuentes}, J.~R. and {Barik}, Ankit and {Fuller}, Jim},
        title = "{Excitation of Inertial Modes in 3D Simulations of Rotating Convection in Planets and Stars}",
      journal = {\apj},
         year = 2026,
        month = feb,
       volume = {998},
       number = {1},
          eid = {131},
        pages = {131},
          doi = {10.3847/1538-4357/ae3538},
archivePrefix = {arXiv},
       eprint = {2511.16630},
 primaryClass = {astro-ph.SR},
       adsurl = {https://ui.adsabs.harvard.edu/abs/2026ApJ...998..131F}
}

@ARTICLE{Vorontsov2002,
       author = {{Vorontsov}, S.~V. and {Christensen-Dalsgaard}, J. and {Schou}, J. and {Strakhov}, V.~N. and {Thompson}, M.~J.},
        title = "{Helioseismic Measurement of Solar Torsional Oscillations}",
      journal = {Science},
         year = 2002,
        month = apr,
       volume = {296},
       number = {5565},
        pages = {101-103},
          doi = {10.1126/science.1069190},
       adsurl = {https://ui.adsabs.harvard.edu/abs/2002Sci...296..101V}
}

@ARTICLE{Loeptien2018,
       author = {{L{\"o}ptien}, Bj{\"o}rn and {Gizon}, Laurent and {Birch}, Aaron C. and {Schou}, Jesper and {Proxauf}, Bastian and {Duvall}, Thomas L. and {Bogart}, Richard S. and {Christensen}, Ulrich R.},
        title = "{Global-scale equatorial Rossby waves as an essential component of solar internal dynamics}",
      journal = {Nature Astronomy},
         year = 2018,
        month = may,
       volume = {2},
        pages = {568-573},
          doi = {10.1038/s41550-018-0460-x},
archivePrefix = {arXiv},
       eprint = {1805.07244},
 primaryClass = {astro-ph.SR},
       adsurl = {https://ui.adsabs.harvard.edu/abs/2018NatAs...2..568L}
}

@ARTICLE{Proxauf2020,
       author = {{Proxauf}, B. and {Gizon}, L. and {L{\"o}ptien}, B. and {Schou}, J. and {Birch}, A.~C. and {Bogart}, R.~S.},
        title = "{Exploring the latitude and depth dependence of solar Rossby waves using ring-diagram analysis}",
      journal = {\aap},
         year = 2020,
        month = feb,
       volume = {634},
          eid = {A44},
        pages = {A44},
          doi = {10.1051/0004-6361/201937007},
archivePrefix = {arXiv},
       eprint = {1912.02056},
 primaryClass = {astro-ph.SR},
       adsurl = {https://ui.adsabs.harvard.edu/abs/2020A&A...634A..44P}
}

@ARTICLE{Hanson2022,
       author = {{Hanson}, Chris S. and {Hanasoge}, Shravan and {Sreenivasan}, Katepalli R.},
        title = "{Discovery of high-frequency retrograde vorticity waves in the Sun}",
      journal = {Nature Astronomy},
         year = 2022,
        month = mar,
       volume = {6},
        pages = {708-714},
          doi = {10.1038/s41550-022-01632-z},
       adsurl = {https://ui.adsabs.harvard.edu/abs/2022NatAs...6..708H}
}

@INPROCEEDINGS{Hill2003,
       author = {{Hill}, Frank and {Bolding}, John and {Toner}, Clifford and {Corbard}, Thierry and {Wampler}, Steve and {Goodrich}, Bret and {Goodrich}, Jean and {Eliason}, Patricia and {Hanna}, Kerri Donaldson},
        title = "{The GONG++ data processing pipeline}",
    booktitle = {GONG+ 2002. Local and Global Helioseismology: the Present and Future},
         year = 2003,
       editor = {{Sawaya-Lacoste}, Huguette},
       series = {ESA Special Publication},
       volume = {517},
        month = feb,
        pages = {295-298},
       adsurl = {https://ui.adsabs.harvard.edu/abs/2003ESASP.517..295H}
}

@ARTICLE{Liang2019,
       author = {{Liang}, Zhi-Chao and {Gizon}, Laurent and {Birch}, Aaron C. and {Duvall}, Thomas L.},
        title = "{Time-distance helioseismology of solar Rossby waves}",
      journal = {\aap},
         year = 2019,
        month = jun,
       volume = {626},
          eid = {A3},
        pages = {A3},
          doi = {10.1051/0004-6361/201834849},
archivePrefix = {arXiv},
       eprint = {1812.07413},
 primaryClass = {astro-ph.SR},
       adsurl = {https://ui.adsabs.harvard.edu/abs/2019A&A...626A...3L}
}

@ARTICLE{Goddard2020,
       author = {{Goddard}, C.~R. and {Birch}, A.~C. and {Fournier}, D. and {Gizon}, L.},
        title = "{Predicting frequency changes of global-scale solar Rossby modes due to solar cycle changes in internal rotation}",
      journal = {\aap},
         year = 2020,
        month = aug,
       volume = {640},
          eid = {L10},
        pages = {L10},
          doi = {10.1051/0004-6361/202038539},
archivePrefix = {arXiv},
       eprint = {2007.14387},
 primaryClass = {astro-ph.SR},
       adsurl = {https://ui.adsabs.harvard.edu/abs/2020A&A...640L..10G}
}

@ARTICLE{Gizon2021,
       author = {{Gizon}, Laurent and {Cameron}, Robert H. and {Bekki}, Yuto and {Birch}, Aaron C. and {Bogart}, Richard S. and {Brun}, Allan Sacha and {Damiani}, Cilia and {Fournier}, Damien and {Hyest}, Laura and {Jain}, Kiran and {Lekshmi}, B. and {Liang}, Zhi-Chao and {Proxauf}, Bastian},
        title = "{Solar inertial modes: Observations, identification, and diagnostic promise}",
      journal = {\aap},
         year = 2021,
        month = aug,
       volume = {652},
          eid = {L6},
        pages = {L6},
          doi = {10.1051/0004-6361/202141462},
archivePrefix = {arXiv},
       eprint = {2107.09499},
 primaryClass = {astro-ph.SR},
       adsurl = {https://ui.adsabs.harvard.edu/abs/2021A&A...652L...6G}
}

@ARTICLE{Anderson1990,
       author = {{Anderson}, Edwin R. and {Duvall}, Thomas L., Jr. and {Jefferies}, Stuart M.},
        title = "{Modeling of Solar Oscillation Power Spectra}",
      journal = {\apj},
         year = 1990,
        month = dec,
       volume = {364},
        pages = {699},
          doi = {10.1086/169452},
       adsurl = {https://ui.adsabs.harvard.edu/abs/1990ApJ...364..699A}
}

@misc{SILSO_Sunspot_Number,
author = {{Clette}, F. and {Lefèvre}, L.},
title = {SILSO Sunspot Number V2.0},
doi = {10.24414/qnza-ac80},
howpublished = {https://doi.org/10.24414/qnza-ac80},
month = {07},
year = {2015},
note = {Published by WDC SILSO - Royal Observatory of Belgium (ROB)}
}

@ARTICLE{Philidet2023,
       author = {{Philidet}, J. and {Gizon}, L.},
        title = "{Interaction of solar inertial modes with turbulent convection. A 2D model for the excitation of linearly stable modes}",
      journal = {\aap},
         year = 2023,
        month = may,
       volume = {673},
          eid = {A124},
        pages = {A124},
          doi = {10.1051/0004-6361/202245666},
archivePrefix = {arXiv},
       eprint = {2304.05926},
 primaryClass = {astro-ph.SR},
       adsurl = {https://ui.adsabs.harvard.edu/abs/2023A&A...673A.124P}
}

@ARTICLE{Snodgrass1984,
       author = {{Snodgrass}, H.~B.},
        title = "{Separation of large-scale photospheric Doppler patterns}",
      journal = {\solphys},
         year = 1984,
        month = aug,
       volume = {94},
       number = {1},
        pages = {13-31},
          doi = {10.1007/BF00154804},
       adsurl = {https://ui.adsabs.harvard.edu/abs/1984SoPh...94...13S}
}

@ARTICLE{Haber2002,
       author = {{Haber}, Deborah A. and {Hindman}, Bradley W. and {Toomre}, Juri and {Bogart}, Richard S. and {Larsen}, Rasmus M. and {Hill}, Frank},
        title = "{Evolving Submerged Meridional Circulation Cells within the Upper Convection Zone Revealed by Ring-Diagram Analysis}",
      journal = {\apj},
         year = 2002,
        month = may,
       volume = {570},
       number = {2},
        pages = {855-864},
          doi = {10.1086/339631},
       adsurl = {https://ui.adsabs.harvard.edu/abs/2002ApJ...570..855H}
}

@article{Bogart2011a,
	author = {{Bogart}, R.~S. and {Baldner}, C. and {Basu}, S. and {Haber}, D.~A. and {Rabello-Soares}, M.~C.},
	doi = {10.1088/1742-6596/271/1/012008},
	url = {https://doi.org/10.1088/1742-6596/271/1/012008},
	year = 2011,
	month = {jan},
	publisher = {{IOP} Publishing},
	volume = {271},
	pages = {012008},
    journal = {Journal of Physics: Conference Series},
	title = {{HMI} ring diagram analysis I. The processing pipeline},
}

@INPROCEEDINGS{Bogart2011b,
       author = {{Bogart}, R.~S. and {Baldner}, C. and {Basu}, S. and {Haber}, D.~A. and {Rabello-Soares}, M.~C.},
        title = "{HMI ring diagram analysis II. Data products}",
    booktitle = {GONG-SoHO 24: A New Era of Seismology of the Sun and Solar-Like Stars},
         year = 2011,
       series = {Journal of Physics Conference Series},
       volume = {271},
        month = jan,
          eid = {012009},
        pages = {012009},
          doi = {10.1088/1742-6596/271/1/012009},
       adsurl = {https://ui.adsabs.harvard.edu/abs/2011JPhCS.271a2009B}
}

@INPROCEEDINGS{Corbard2003,
       author = {{Corbard}, T. and {Toner}, C. and {Hill}, F. and {Hanna}, K.~D. and {Haber}, D.~A. and {Hindman}, B.~W. and {Bogart}, R.~S.},
        title = "{Ring-diagram analysis with GONG++}",
    booktitle = {GONG+ 2002. Local and Global Helioseismology: the Present and Future},
         year = 2003,
       editor = {{Sawaya-Lacoste}, Huguette},
       series = {ESA Special Publication},
       volume = {517},
        month = feb,
        pages = {255-258},
       adsurl = {https://ui.adsabs.harvard.edu/abs/2003ESASP.517..255C}
}

@ARTICLE{Bekki2022nonlinear,
       author = {{Bekki}, Yuto and {Cameron}, Robert H. and {Gizon}, Laurent},
        title = "{Theory of solar oscillations in the inertial frequency range: Amplitudes of equatorial modes from a nonlinear rotating convection simulation}",
      journal = {\aap},
         year = 2022,
        month = oct,
       volume = {666},
          eid = {A135},
        pages = {A135},
          doi = {10.1051/0004-6361/202244150},
archivePrefix = {arXiv},
       eprint = {2208.11081},
 primaryClass = {astro-ph.SR},
       adsurl = {https://ui.adsabs.harvard.edu/abs/2022A&A...666A.135B}
}

@ARTICLE{Bekki2022linear,
       author = {{Bekki}, Yuto and {Cameron}, Robert H. and {Gizon}, Laurent},
        title = "{Theory of solar oscillations in the inertial frequency range: Linear modes of the convection zone}",
      journal = {\aap},
         year = 2022,
        month = jun,
       volume = {662},
          eid = {A16},
        pages = {A16},
          doi = {10.1051/0004-6361/202243164},
archivePrefix = {arXiv},
       eprint = {2203.04442},
 primaryClass = {astro-ph.SR},
       adsurl = {https://ui.adsabs.harvard.edu/abs/2022A&A...662A..16B}
}

@article{Waidele2023,
doi = {10.3847/2041-8213/acefd0},
url = {https://dx.doi.org/10.3847/2041-8213/acefd0},
year = {2023},
month = {sep},
publisher = {The American Astronomical Society},
volume = {954},
number = {1},
pages = {L26},
author = {M. Waidele and Junwei Zhao},
title = {Observed Power and Frequency Variations of Solar Rossby Waves with Solar Cycles},
journal = {\apjl}
}

@ARTICLE{Liang2025,
       author = {{Liang}, Zhi-Chao and {Gizon}, Laurent},
        title = "{Doppler velocity of m = 1 high-latitude inertial mode over the last five sunspot cycles}",
      journal = {\aap},
         year = 2025,
        month = mar,
       volume = {695},
          eid = {A67},
        pages = {A67},
          doi = {10.1051/0004-6361/202452133},
archivePrefix = {arXiv},
       eprint = {2409.06896},
 primaryClass = {astro-ph.SR},
       adsurl = {https://ui.adsabs.harvard.edu/abs/2025A&A...695A..67L}
}

@INPROCEEDINGS{Gizon2024,
       author = {{Gizon}, Laurent and {Bekki}, Yuto and {Birch}, Aaron C. and {Cameron}, Robert H. and {Fournier}, Damien and {Philidet}, Jordan and {Lekshmi}, B. and {Liang}, Zhi-Chao},
        title = "{Solar Inertial Modes}",
    booktitle = {IAU Symposium},
         year = 2024,
       editor = {{Getling}, Alexander V. and {Kitchatinov}, Leonid L.},
       series = {IAU Symposium},
       volume = {365},
        month = dec,
        pages = {207-221},
          doi = {10.1017/S1743921324000425},
       adsurl = {https://ui.adsabs.harvard.edu/abs/2024IAUS..365..207G}
}

@ARTICLE{Bekki2024,
       author = {{Bekki}, Yuto and {Cameron}, Robert H. and {Gizon}, Laurent},
        title = "{The Sun's differential rotation is controlled by high-latitude baroclinically unstable inertial modes}",
      journal = {Science Advances},
         year = 2024,
        month = mar,
       volume = {10},
       number = {13},
          eid = {eadk5643},
        pages = {eadk5643},
          doi = {10.1126/sciadv.adk5643},
archivePrefix = {arXiv},
       eprint = {2403.18986},
 primaryClass = {astro-ph.SR},
       adsurl = {https://ui.adsabs.harvard.edu/abs/2024SciA...10K5643B}
}

@Article{Scherrer2012,
  author  = {{Scherrer}, P.~H. and {Schou}, J. and {Bush}, R.~I. and {Kosovichev}, A.~G. and {Bogart}, R.~S. and {Hoeksema}, J.~T. and {Liu}, Y. and {Duvall}, T.~L. and {Zhao}, J. and {Title}, A.~M. and {Schrijver}, C.~J. and {Tarbell}, T.~D. and {Tomczyk}, S.},
  title   = {{The Helioseismic and Magnetic Imager (HMI) Investigation for the Solar Dynamics Observatory (SDO)}},
  journal = {\solphys},
  year    = {2012},
  volume  = {275},
  pages   = {207-227},
  month   = jan,
  adsurl  = {http://adsabs.harvard.edu/abs/2012SoPh..275..207S},
  doi     = {10.1007/s11207-011-9834-2},
}

@Article{Schou2012,
  author   = {{Schou}, J. and {Scherrer}, P.~H. and {Bush}, R.~I. and {Wachter}, R. and {Couvidat}, S. and {Rabello-Soares}, M.~C. and {Bogart}, R.~S. and {Hoeksema}, J.~T. and {Liu}, Y. and {Duvall}, T.~L. and {Akin}, D.~J. and {Allard}, B.~A. and {Miles}, J.~W. and {Rairden}, R. and {Shine}, R.~A. and {Tarbell}, T.~D. and {Title}, A.~M. and {Wolfson}, C.~J. and {Elmore}, D.~F. and {Norton}, A.~A. and {Tomczyk}, S.},
  title    = {{Design and Ground Calibration of the Helioseismic and Magnetic Imager (HMI) Instrument on the Solar Dynamics Observatory (SDO)}},
  journal  = {\solphys},
  year     = {2012},
  volume   = {275},
  pages    = {229-259},
  month    = jan,
  adsurl   = {http://adsabs.harvard.edu/abs/2012SoPh..275..229S},
  doi      = {10.1007/s11207-011-9842-2},
}

@ARTICLE{Harvey1996,
       author = {{Harvey}, J.~W. and {Hill}, F. and {Hubbard}, R.~P. and {Kennedy}, J.~R. and {Leibacher}, J.~W. and {Pintar}, J.~A. and {Gilman}, P.~A. and {Noyes}, R.~W. and {Title}, A.~M. and {Toomre}, J. and {Ulrich}, R.~K. and {Bhatnagar}, A. and {Kennewell}, J.~A. and {Marquette}, W. and {Patron}, J. and {Saa}, O. and {Yasukawa}, E.},
        title = "{The Global Oscillation Network Group (GONG) Project}",
      journal = {Science},
         year = 1996,
        month = may,
       volume = {272},
       number = {5266},
        pages = {1284-1286},
          doi = {10.1126/science.272.5266.1284},
       adsurl = {https://ui.adsabs.harvard.edu/abs/1996Sci...272.1284H}
}

@ARTICLE{Howe2009,
       author = {{Howe}, Rachel},
        title = "{Solar Interior Rotation and its Variation}",
      journal = {Living Reviews in Solar Physics},
         year = 2009,
        month = dec,
       volume = {6},
       number = {1},
          eid = {1},
        pages = {1},
          doi = {10.12942/lrsp-2009-1},
archivePrefix = {arXiv},
       eprint = {0902.2406},
 primaryClass = {astro-ph.SR},
       adsurl = {https://ui.adsabs.harvard.edu/abs/2009LRSP....6....1H}
}

\begin{appendix}

\section{Frequency windows for measurements of mode parameters}

\begin{table}[h]
\caption{Frequency windows used for mode fitting and parameter measurements.}
\centering
\renewcommand{\arraystretch}{1.1}
\label{tab:window}
\begin{tabular}{r c c}
\hline \hline
& \multicolumn{2}{c}{Frequency windows [nHz, nHz]} \\
$m$ & mean power spectrum & 4-year power spectra \\
\hline
& \multicolumn{2}{c}{\it{High-latitude mode}} \\
$1$  & $[-190,\,\,  110]$ & $[-118,\,\;\:      -56]$ \\
\hline
& \multicolumn{2}{c}{\it{Equatorial Rossby modes}} \\
$3$  & $[-350,\,    -50]$ & $[-283,\,         -243]$ \\
$4$  & $[-370,\,    -70]$ & $[-245,\,         -173]$ \\
$5$  & $[-370,\,    -70]$ & $[-201,\,         -140]$ \\
$6$  & $[-310,\,    -10]$ & $[-191,\,\;\:      -89]$ \\
$7$  & $[-280,\;\;\; 20]$ & $[-158,\,\;\:      -98]$ \\
$8$  & $[-240,\;\;\; 60]$ & $[-158,\,\;\:      -68]$ \\
$9$  & $[-300,\,\,  100]$ & $[-163,\,\;\:      -43]$ \\
$10$ & $[-310,\;\;\; 90]$ & $[-149,\,\;\:      -48]$ \\
$11$ & $[-300,\,\,  100]$ & $[-146,\,\;\:      -20]$ \\
$12$ & $[-310,\;\;\; 90]$ & $[-145,\,\;\:      -41]$ \\
$13$ & $[-290,\,\,  110]$ & $[-164,\quad\;      28]$ \\
$14$ & $[-290,\,\,  110]$ & $[-142,\quad\;\;\:   1]$ \\
$15$ & $[-290,\,\,  110]$ & $[-133,\quad\:      -4]$ \\
$16$ & $[-290,\,\,  110]$ & $[-135,\quad\;\;\;   8]$ \\
\hline
\end{tabular}
\tablefoot{For each mode, the table lists the frequency window used to fit the GONG reference power spectrum, together with the frequency window $W$ used to measure the mode parameters in each time segment for both GONG and HMI. Frequencies are given in the Carrington frame in nHz.}
\end{table}

\section{Fitting the power spectrum averaged over multiple realizations}
\label{sec:Likelihood_function}

\subsection{Statistics of the average power spectrum}
At fixed frequency $\nu$, we considered $n$ independent realizations of the power spectrum,  $P_i=U_i^2+V_i^2$, where $U_i$ and $V_i$ are independent zero-mean Gaussian random variables satisfying $\mathbb{E}[U_i^2]=\mathbb{E}[V_i^2]=\mathscr{P}/2$.
We denote the average by $\overline{P} = \sum_{i=1}^n P_i /n$.
The random variable $X= 2n \overline{P} / \mathscr{P}$ has an expectation value equal to  $2n$ and is thus described by a $\chi^2$ with $k=2n$ degrees of freedom with probability distribution function (PDF)
\begin{equation}
    f_X(x) = c_k\  x^{k/2 - 1}e^{-x/2}, \quad x \ge 0  ,
\end{equation}
where $c_k$ is a normalization constant.
The PDF of $\overline{P}$ is obtained by the change of variable  $\overline{p}= (\mathscr{P}/2n) x$:
\begin{equation}
f(\overline{p}; \mathscr{P}) =  \frac{2n}{\mathscr{P}} f_X\left( \frac{2n \overline{p}}{\mathscr{P}} \right)
  =
  c_{2n} (2n)^n
  \left( \frac{ \overline{p}^{\, n-1}}{\mathscr{P}^n} \right)  \exp\left( -\frac{n\overline{p}}{\mathscr{P}} \right)
  .
\end{equation}

\subsection{Maximum likelihood estimation}
The  mode parameters are measured from the averaged power spectrum $\overline{p}_j$ over a range of independent frequencies $\nu_j$, with $1\leq j \leq J$. The model for the expectation value of the power spectrum is denoted by  $\mathscr{P}(\nu_j,\vec{q})$,
where the parameters $\vec{q}$ are to be determined.
The likelihood  is given by the joint PDF evaluated at the sample data:
\begin{equation}
    \mathcal{L} (\vec{q})= \prod_{j=1}^J f\left(\overline{p}_j; \mathscr{P} (\nu_j,\vec{q})\right) .
\end{equation}
The parameters $\vec{q}$ are inferred by maximizing the likelihood function evaluated at the sample data, or, equivalently, minimizing
\begin{eqnarray}
    -\ln \mathcal{L} (\vec{q})  & =  &    n \sum_{j=1}^J \left( \ln \mathscr{P}(\nu_j, \vec{q}) + \frac{\overline{p}_j}{\mathscr{P}(\nu_j, \vec{q} )}\right)   \nonumber \\
    && +
    \textrm{ terms independent of $\vec{q}$} .
\end{eqnarray}
 Hence, the maximum likelihood parameters are
 \begin{equation}
 \vec{q}^\ast = \textrm{arg} \min_{\vec{q}}  \sum_{j=1}^J \left( \ln \mathscr{P}(\nu_j, \vec{q}) + \frac{\overline{p}_j}{\mathscr{P}(\nu_j, \vec{q} )}\right) .  \end{equation}
 We conclude that the estimated parameters  are obtained
 by fitting the averaged power spectrum as if it were a single realization of the power spectrum \citep[see][]{Anderson1990}.

\section{Additional figures}

\begin{figure}[h]
\centering
    \includegraphics[width=\hsize]{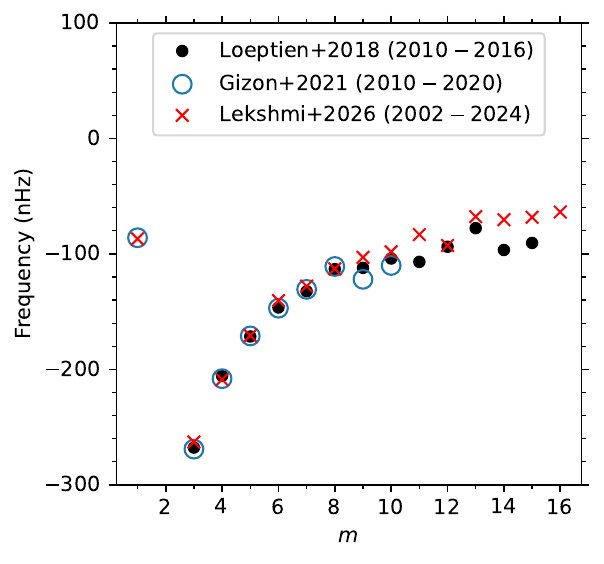}
    \caption{Comparison of mode frequencies from \citet[filled black circles; HMI 2010--2016]{Loeptien2018}, \citet[open blue circles; HMI 2010--2020]{Gizon2021}, and Table~\ref{tab:parameters} (red crosses; GONG 2002--2024). The differences primarily reflect the differing time spans of the datasets.  The scatter increases with $m$, consistent with the stronger solar-cycle variability shown in Fig.~\ref{fig:power_freq_ssn}b.
    } \label{fig:cf-freq}
\end{figure}

\begin{figure*}[h]
    \centering
    \includegraphics[width=\textwidth]{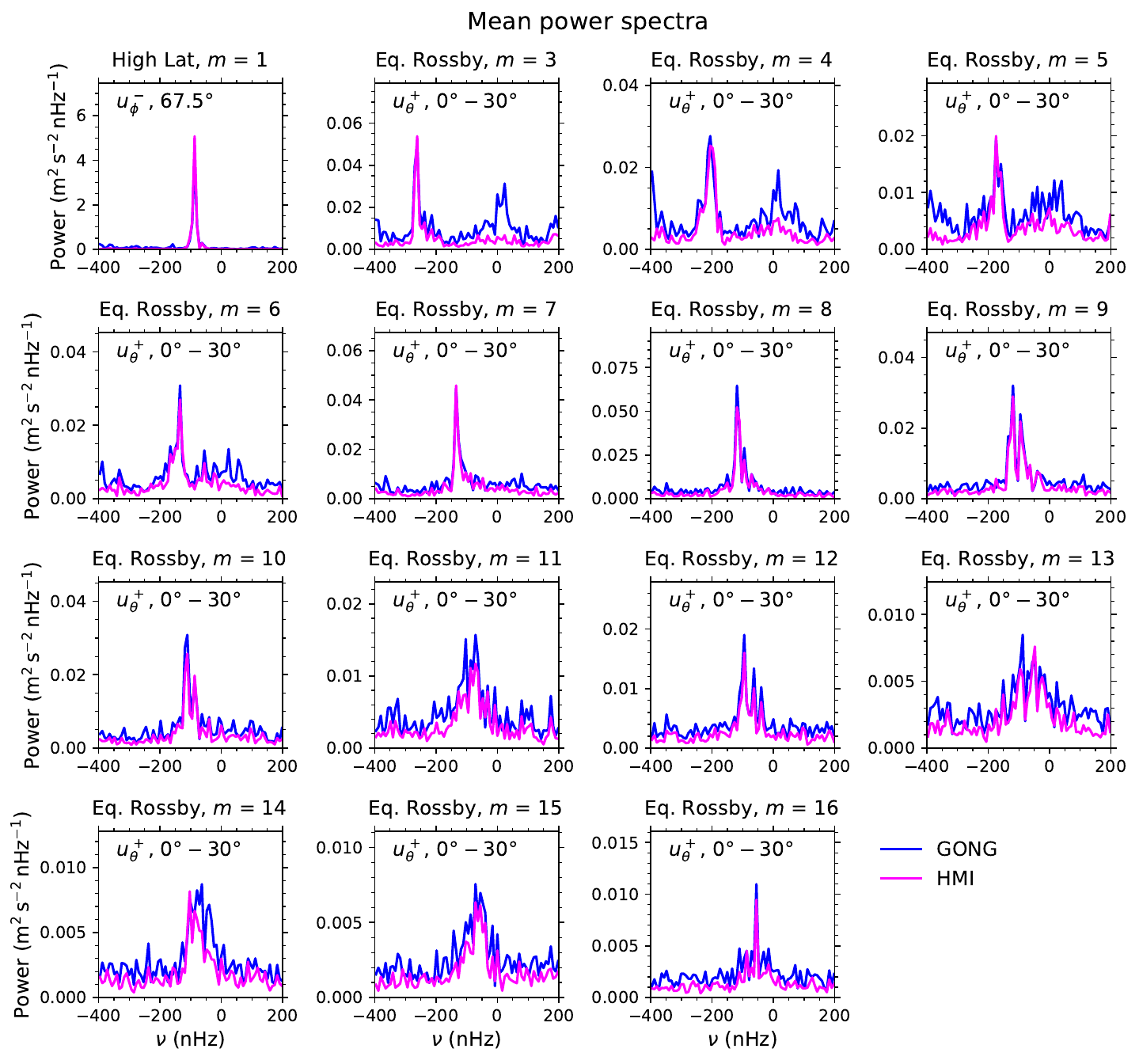}
    \caption{Mean power spectra $P_m(\nu)$ of $u_{\phi}^-$ at $m = 1$ and of $u_{\theta}^+$ at $3\le m \le 16,$ showing the HL and ER modes, respectively. They are computed from HMI (magenta) and GONG (blue) data for the period 2010.5--2022.5.}
    \label{fig:meanspectra_HMI_GONG}
\end{figure*}

\begin{figure*}[h]
    \centering
    \includegraphics[width=\textwidth]{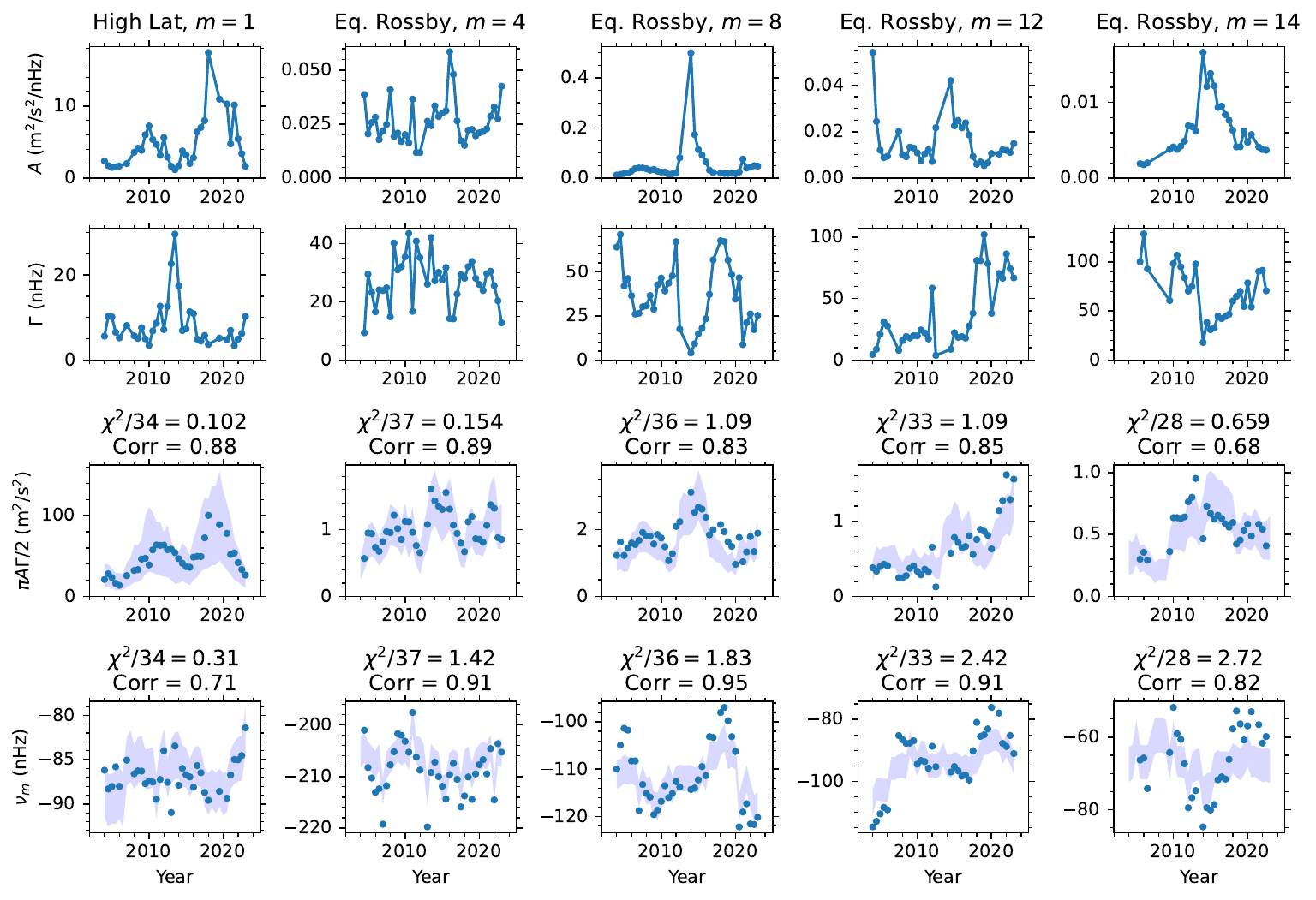}
    \caption{Mode parameters (including linewidths) for the $m=1$ HL mode and $m=4$, 8, 12, and 14 ER modes from Lorentzian fits.
    The dots in rows 1, 2, and 4 show the heights ($A_m$), full widths at half maximum ($\Gamma_m$), and mode frequencies ($\nu_m$) obtained for each 4-year time segment by fitting Lorentzian profiles to the data. The dots in the third row show the corresponding products, $\pi A_m \Gamma_m / 2$, which provide a measure of the mode power inferred from the fits.
    In rows 3 and 4, shaded areas are included to illustrate the consistency with measurements obtained using the more robust method described in Sect.~\ref{sec:parameter_calc} (see Eqs.~\eqref{eq:frequency_calc} and \eqref{eq:power_calc}). The reduced $\chi^2$ values quantifying the differences between the two methods, as well as the corresponding correlation coefficients, are indicated.
    }
    \label{fig:lorentzianfit}
\end{figure*}

\begin{figure*}
    \centering
    \includegraphics[width=\textwidth]{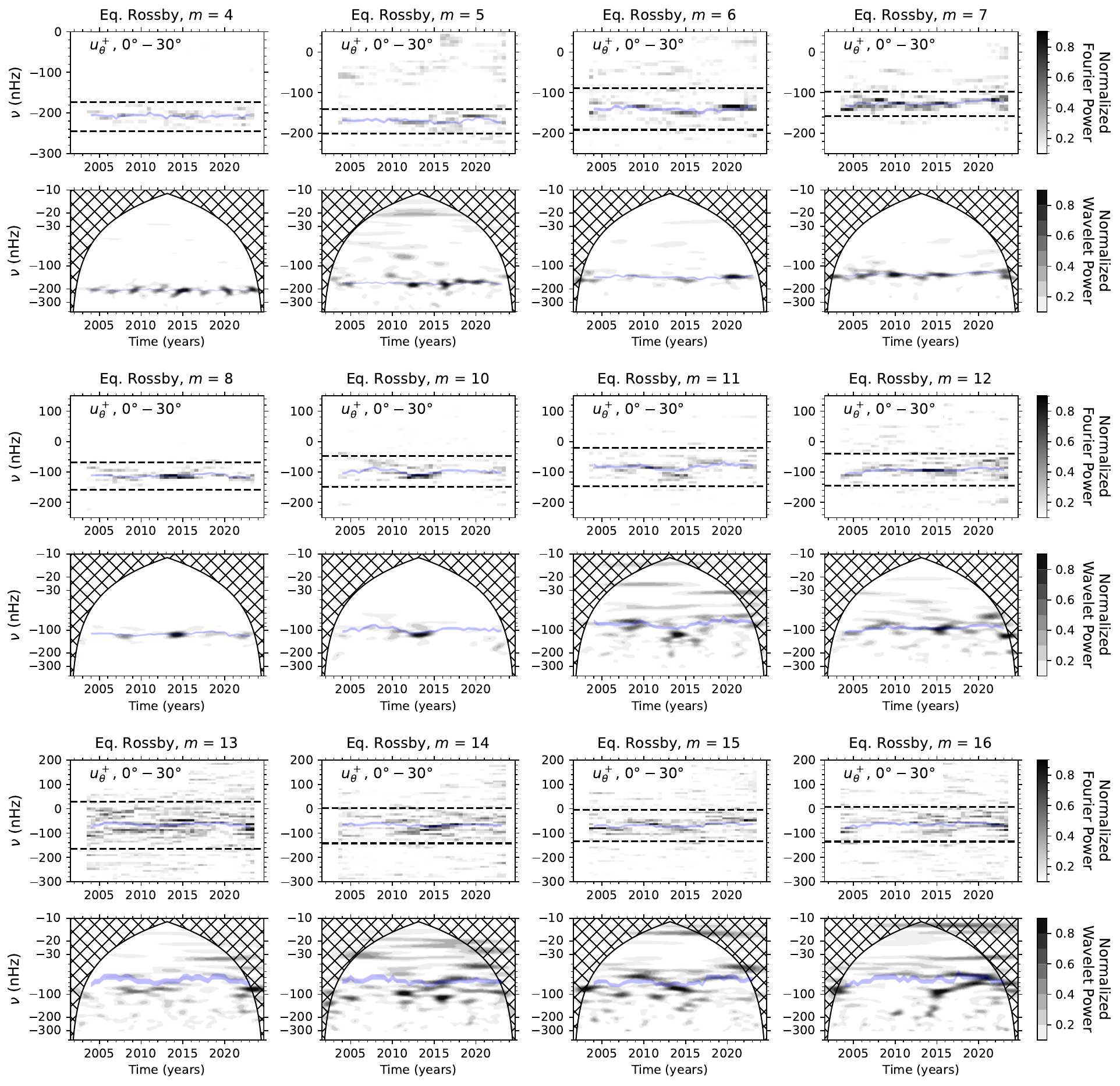}
    \caption{Same as Fig.~\ref{fig:2D_spectra} but for the remaining modes.}
    \label{fig:2D_spectra_allmodes}
\end{figure*}

\end{appendix}
\end{document}